\documentclass[a4paper,11pt]{article}

\usepackage{jinstpub} 
\usepackage{lineno}
\usepackage{hyperref}
\usepackage[figure]{hypcap}
\usepackage{placeins}

\title{\boldmath Development, Characterization and Production of a novel Water-based Liquid Scintillator based on the Surfactant TRITON\texttrademark{} X-100}

\author[a,b,c,1]{Hans Th. J. Steiger,\note{Corresponding author.}}
\author[b,c]{Manuel Böhles,}
\author[a]{Matthias Raphael Stock,}
\author[b,c]{Michael Wurm,}
\author[a]{David Dörflinger,}
\author[a]{Ulrike Fahrendholz,}
\author[b]{Anastasia Mpoukouvalas,}
\author[a]{Lothar Oberauer,}
\author[a]{Andreas Steiger,}
\author[b,c]{Dorina Zundel}

\affiliation[a]{Technical University of Munich, TUM School of Natural Sciences, Physics Department, \\ James-Franck-Str. 1, 85748 Garching, Germany}%{Physik-Department E15, Technische Universität München \\ James-Franck-Straße 1, 85748 Garching, Germany}%Technical University of Munich,\\James Franck Straße 1, 85748 Garching, Germany}
\affiliation[b]{Cluster of Excellence PRISMA$^+$ \\Staudingerweg 9, 55128 Mainz, Germany}
\affiliation[c]{Institute for Physics, Johannes Gutenberg University Mainz \\ Staudingerweg 7, 55128 Mainz, Germany}

\emailAdd{hans.steiger@tum.de}

\abstract{Water-based Liquid Scintillator~(WbLS) is a novel detector medium for particle physics experiments. Applications range from the use as hybrid Cherenkov/scintillation target in low-energy and accelerator neutrino experiments to large-volume neutron vetoes for dark matter detectors. Here, we present a novel WbLS featuring new components (the surfactant Triton-X and vitamin~C for long-term stability), a new production recipe, and a thorough characterization of its properties. Moreover, based on neutron scattering data we are able to demonstrate that the pulse shape discrimination capabilities of this particular~LS are comparable to fully-organic~LAB based scintillators.}


\arxivnumber{1234.56789} % only if you have one

\begin{document}
\maketitle
\flushbottom

\section{Introduction}

Water-based Liquid Scintillator~(WbLS) is a novel detection medium for large-volume neutrino detectors. The development aims at the observation of neutrinos in the~MeV to~GeV energy range,~i.e.~covering a wide range of astrophysical (solar and Supernova neutrinos) and terrestrial sources~(reactor, geo- and atmospheric neutrinos) as well as long-baseline oscillation experiments using accelerator neutrinos \cite{Theia:2019non}. Especially if loaded with gadolinium,~WbLS is an attractive option for use as active neutron veto for next-generation dark matter searches \cite{wang:2023}. At the time of writing, with~ANNIE, \textsc{Eos},~BUTTON and the~BNL prototypes several ton-scale demonstrators have either taken first data or a close to operation, investigating different aspects of the applicability of~WbLS to accelerator or low-energy neutrinos, radiopurity as well as purification and long-term chemical stability \cite{ANNIE:2023yny,Anderson:2022lbb,Akindele:2023ixz,Zhao:2023ydx}.

WbLS properties are optimized to permit a hybrid detection of the Cherenkov and scintillation photons that are emitted by the final state leptons and hadrons in neutrino interactions \cite{ASDC:2014}. The relatively low organic fraction results in a high transparency in the near-UV/blue range of the spectrum where most of the scintillation and Cherenkov light transport occurs. Most~WbLS mixtures feature inherently fast scintillation emission times (on the nanosecond scale) and a light yield roughly proportional to the contained organic fraction. In the~MeV range, these properties can be exploited to separately reconstruct and compare the Cherenkov and scintillation light output of the final state particle(s), offering enhanced particle discrimination based on differing Cherenkov/scintillation ratios~(e.g.~\cite{Sawatzki:2020mpb,Zsoldos:2022mre}). At~GeV energies,~WbLS is expected to feature improved vertex resolution based on the scintillation signal of recoil hadrons. This enhances the energy reconstruction of final-state muons but in addition provides an energy estimate for hadronic recoils, overall improving the reconstruction of neutrino energy. A baseline concept for a~WbLS-based neutrino detector \textsc{Theia} on the mass scale of 25--100 kilotons has been laid out in Ref.~\cite{Theia:2019non}.

In the past, several laboratory studies have been undertaken to determine the properties of WbLS mixtures based on the organic solvent linear alkyl-benzene~(LAB) and the surfactant linear alkyl-sulfonate~(LAS). Here, we report results from the characterization of a novel WbLS combining LAB with the common surfactant Triton-X. Compared to~LAS, Triton-X features the advantages of larger optical transparency and self-scintillation. Moreover, the surfactant is non-ionic and known for its high solubility in water. The resulting~Triton-X based~WbLS offers similar properties to earlier LAS-based variants~\cite{Kaptanoglu:2021prv,Callaghan:2022ahi} and at the same time proofs easier to prepare in liter-scale batches. Moreover, we add vitamin C at small concentration to improve the chemical stability of the scintillator, especially concerning oxidation of the~PPO.

The paper is structured as follows: In \autoref{sec:production}, we describe the material composition and production method for the new~Triton-X WbLS. We continue to present its wavelength-dependent transparency, comparing to an estimate of the Rayleigh scattering length estimated based on the average micelle diameter and concentration in \autoref{sec:transparency}. In \autoref{sec:scintillation}, we report the results of laboratory studies determining the light yield as well as the fluorescence spectra and time distribution of the new WbLS. We complement this by measurements of the scintillation pulse shapes acquired during an irradiation campaign at the~LNL accelerator lab in Legnaro, studying the discrimination capability of gamma and neutron signals and comparing to standard organic scintillators. We summarize our findings in \autoref{sec:conclusions}.

\section{Composition and production}
\label{sec:production}
The~components of the new WbLS are listed in \autoref{tab:chems}. The bulk material of the new WbLS is~86\%~vol.~of distilled water, produced by the  in-house plant of the Institute of Physics in Mainz. Typically, this water has an attenuation length of more than~$\sim$2.5\,m for light of~430\,nm wavelength. In the first production step the water is mixed with 25\,mg/l of ~L(+)-Ascorbic acid as antioxidant. After that,~13\%vol.~of the surfactant 2-[4-(2,4,4-trimethylpentan-2-yl)phenoxy]ethanol, better known as Triton~X-100, of purissimum grade is slowly dissolved in the water while continuously stirring. During this process the temperature of the solution is kept constant at~20$^\circ$C - 24$^\circ$C. Immediately afterwards, the solution is homogenized for~10 minutes using an ultrasonic homogenizer~(Bandelin, Sonopuls HD4400) booster horn. In order to keep the heating of the solution small and the liquid's temperature within the mentioned interval, this step takes place on a bath of ice water.        

\begin{table}[h]
\centering
\begin{tabular}{|c|c|c|c|}
\hline 
Substance & CAS Number & Function & Concentration \\ 
\hline 
Distilled Water & 7732-18-5 & main solvent & 86\% vol \\
\hline 
Alkylphenylpolyethylenglykol & 9002-93-1 & surfactant & 13\% vol \\ (Triton X-100) &  & & \\ 
\hline 
Linear alkyl benzene (LAB) &  & organic solvent & 1\% vol \\ 
\hline 
2,5-Diphenyloxazole & 92-71-7 & oragnic fluor & 100\,g/l (in organic LS) \\ & & & $\sim$1\,g/l (in WbLS) \\
\hline 
L(+)-Ascorbic acid & 50-81-7 & anti-oxidant & 25\,mg/l (in WbLS) \\
\hline 
\end{tabular} 
\caption{Shown is the composition of the~WbLS. The main component is an aqueous solution of Triton~X-100. To protect the scintillation cocktail from oxidation also~25\,mg/l Ascorbic acid are added to the water. The organic phase consists of LAB mixed with the fluor~PPO.}
\label{tab:chems}
\end{table}

In parallel, the organic components (1\%~vol.) are prepared. The linear alkyl benzene~(LAB) of scintillation grade purity was purchased from~HELM~AG. Before admixture to the WbLS, the LAB was further purified by the use of a~800\,mm long chromatography column filled with basic~Al$_2$O$_3$ powder~(CAS: 1344-28-1) in the activity stage~I~(0\% water) to enhance optical transparency. After the chromatography and filtration by means of a Büchner funnel equipped with ash-free filter paper (retention: 2\,\textmu m), the~LAB solvent reaches an attenuation length of several meters for light with a wavelength of~350\,nm and longer. The fluor~2,5-Diphenyloxazole~(PPO) of scintillation grade~(supplied by Carl Roth) was extensively pre-dryed for one week in an evacuated desiccator over a silica gel pearls drying agent. The dry fluor was subsequently dissolved in the~LAB in a concentration of~100\,g/l. To aid this process, the solution was carefully heated to~50$^\circ$C and stirred under nitrogen atmosphere for approx.~3\,h. To rule out the presence of crystalline undissolved~PPO in the organic scintillator that could later interfere with~WbLS synthesis, the organic LS was passed through a~2\,\textmu m filter paper. After this step, the LS was again saturated with dried nitrogen using a gas wash bottle with a filter plate~(porosity 100\,\textmu m - 160\,\textmu m).

In the actual synthesis process (i.e.~filling the micelles with scintillator), the~LAB/PPO mixture was slowly added drop-wise to the aqueous~TX solution using a burette with constant slow stirring. This was necessary to avoid a milky cloudiness of the solution and to ensure that the~LAB and PPO molecules are introduced into the micelles as completely as possible. Subsequently, the mixture was stirred slowly for another~60\,min and then homogenized again for~$\sim$15\,min (stabilized temperature:~$\sim$~20$^\circ$C) with the ultrasonic booster horn.

In the last production step, the~WbLS was pressed through syringe filters with~0.2\,\textmu m pore size to remove agglomerates of micelles and Triton-X~(if present). Subsequently, the~WbLS was carefully saturated with nitrogen by extensively slow bubbling and stored under a protective nitrogen atmosphere.

\section{Transparency and mechanical properties}
\label{sec:transparency}

\begin{figure}[h]
    \centering
    \includegraphics[width=0.6\textwidth]{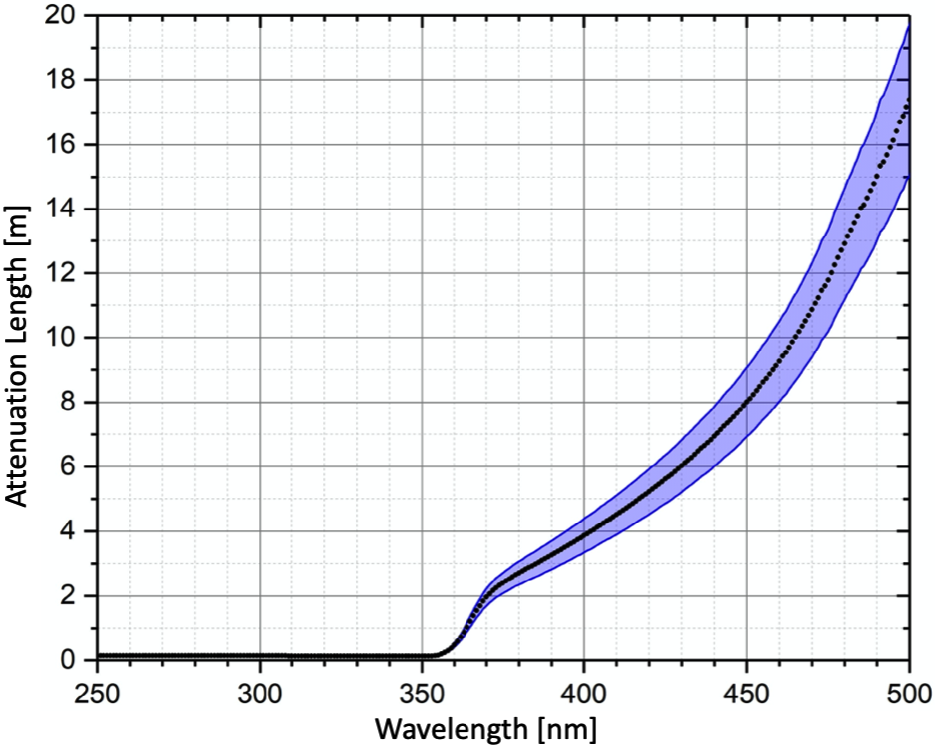}
    \caption{Spectral Attenuation Length of the WbLS. The dotted black curve represents the mean out of 10 measurements, while the blue band indicates their standard deviation as statistical uncertainty. For reasons of clarity, the systematic error has not been presented. }
    \label{fig:AttLength}
\end{figure}

In a first step, we evaluate the optical transparency and mechanical properties of the WbLS which are the main indicators for a successful creation of the suspension. The WbLS transparency is dominated by Rayleigh scattering (Tyndall effect) of the micelles enclosing the organic component. Generally speaking, the generation of small nanometer-scale micelles is reducing the Rayleigh scattering cross section per unit volume and thus the corresponding light attenuation. Here, we investigated both quantities for consistency.

\paragraph{Light attenuation:}
In order to evaluate the transparency of the WbLS the attenuation length as a function of wavelength was studied with a Perkin Elmer Lambda~850+~UV/Vis-spectrometer equipped with a long rectangular fused-silica cuvette (optical path length of 10\,cm). The obtained results are shown in \autoref{fig:AttLength}. The attenuation lengths at $\lambda = 365$\,nm, corresponding to the emission maximum of WbLS, and at $\lambda = 430$\,nm, where conventional phototubes show their maximum spectral quantum efficiency, are found to be 
\begin{eqnarray}
\Lambda_{\text{WbLS}}(365\,\text{nm}) & = & 1.21\, \text{m} \pm 0.16\, \text{m} \,(\text{stat.}) \pm 0.23\, \text{m} \,(\text{syst.}). \nonumber\\
\Lambda_{\text{WbLS}}(430\,\text{nm}) & = & 6.02\, \text{m} \pm 0.81\, \text{m} \,(\text{stat.}) \pm 1.13\, \text{m} \,(\text{syst.}).\nonumber
\end{eqnarray}
While the statistical uncertainty is defined by the variation in between repeated measurements of the same WbLS sample, the systematic uncertainty arises from an empirical correction for reflective losses at the entry and exit of the light beam at the outer glass surfaces of the cuvette.

\paragraph{Viscosity:} The viscosity of the~WbLS sample is an interesting quantity for the handling of the liquid in large-scale experiments and is an important auxiliary quantity for the determination of the micelle size by dynamic light scattering (see below). It has been assessed by a rotational viscometer (Anton Paar, ViscoQC 300L) equipped with an ultralow viscosity spindle~(UL26). We measure the dynamic viscosity as a function of shear stress and obtain a value of~$\eta=(2.86\pm0.14)$\,mPa$\cdot$s at room temperature. We see no signs of a non-Newtonian behavior of the viscosity down to a shear stress of~0.25\,N/mm$^2$, corresponding to the sensitivity limit of the instrument that we have determined using a low-viscosity calibration oil.

\begin{figure}[h]
    \centering
    \includegraphics[width=0.6\textwidth]{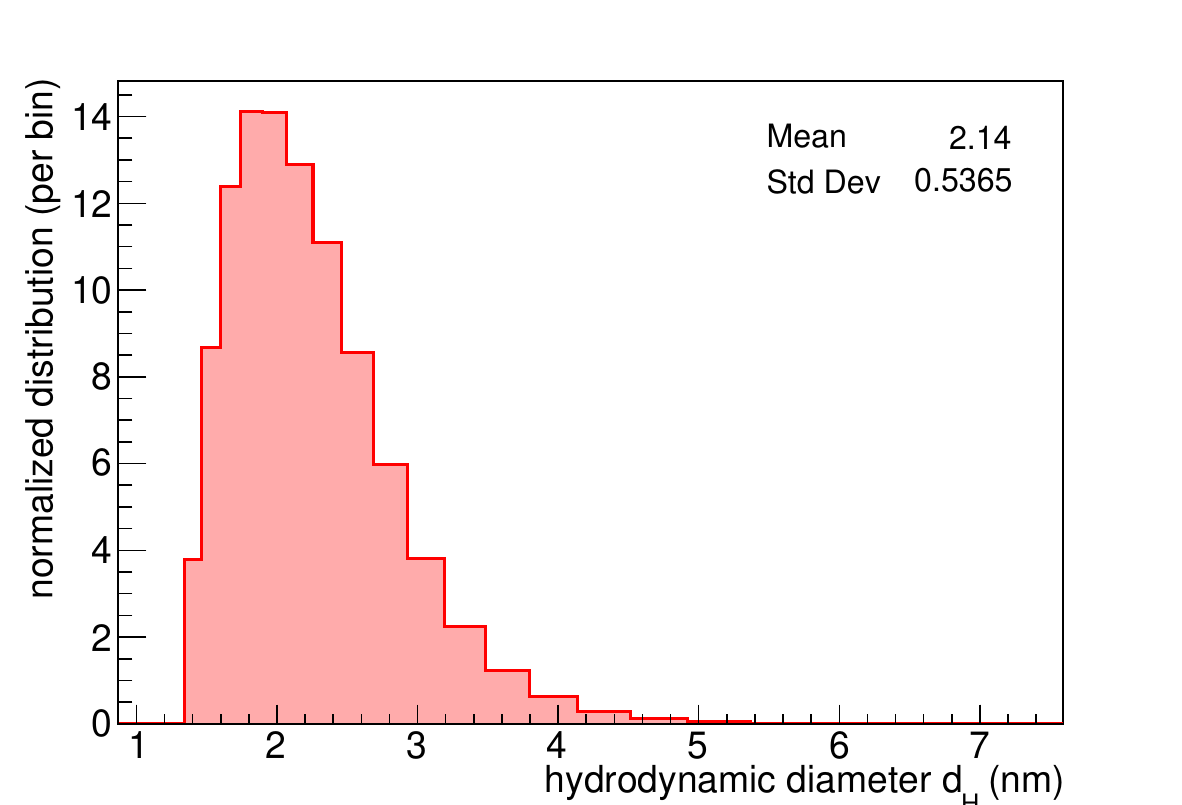}
    \caption{Distribution of hydrodynamic diameters of the~WbLS micelles as determined by a~DLS measurement.}
    \label{fig:dls}
\end{figure}

\paragraph{Micelle size:} We determined the diameter of the micelles using Dynamic Light Scattering (DLS). The tabletop device~(NANO-flex II, Colloid Metrix) is sensitive in the range from~0.3\,nm to~10\,\textmu{}m for minimum concentrations on the ppm level. In the~DLS measurement, the hydrodynamic diameter $d_H$ results from the relation
\begin{equation}
    d_H = \frac{k_BT}{6\pi\eta D},
\end{equation}
where the time correlation of the scattered intensity is used to determine the diffusion coefficient~$D$ of the micelles at room temperature~$T=20^\circ$C, and correct by the viscosity~$\eta$ to obtain the micelle size. \autoref{fig:dls} shows the scattered light intensity as a function of the derived micelle diameter. The distribution peaks at~2.8\,nm with an~RMS of~0.5\,nm. Assuming a density close to water for the micelles (corresponding to an average of~LAB and~TX densities), the derived average mass per micelle is~$\sim$6\,kDa or about~20\,$-$\,25 organic molecules. 

Based on the initial concentration of solvent and surfactant (14\%\,vol.) and the average geometric size of the micelles $\langle V \rangle = 11.5\,{\rm nm}^3$, we determine the size-averaged micelle number density to $\langle n \rangle = 2.4\times10^{24}\,{\rm m}^{-3}$. In addition, we compute the forward scattering scattering cross section per individual micelle, $\langle b^2(0)\rangle = (\langle V_m\rangle \Delta n)^2/\lambda^4 = 8\times 10^{-10}\,{\rm nm}^2$ at $\lambda=430\,{\rm nm}$ \cite{refId0}, with $\Delta n=1.49-1.33=0.16$ the difference in refractive indeces of LAB/Triton-X with respect to water. Based on these quantities, we estimate the Rayleigh scattering length in the visible spectrum to be order of tens of meters, suggesting that the measured attenuation length is dominated by the initial quality of the distilled water.

%We estimate the corresponding scattering length based on Ref.~\ref{}: We first adopt the expression for the total scattered intensity 
%\begin{equation}
%    I(q) = I_0nb^2(0)P(q)S(q)
%\end{equation}
%\textcolor{red}{[Note: One can use this information and the Rayleigh cross-section to calculate the expected scattering length which is on the order of 1.2\,m at 400\,nm -- so about a factor 3 shorter than UV-Vis measurements. The only easy way out would be to assume a smaller organic fraction than 0.13. So I'm not sure we want to include this.]}

%\floatbarrier

\section{Scintillation properties}
\label{sec:scintillation}

In this section, we describe the properties of the new WbLS essential for scintillation light emission: The number of photons emitted per unit deposited energy of a charged particle (light yield, \autoref{sec:ly}), the wavelength spectrum of scintillation (\autoref{sec:es}) and the time profile of light emission (\autoref{sec:fluorescence}). While the first quantity has been determined by a customized laboratory setup using Compton-scattering of gamma rays for excitation of the WbLS, a spectro-fluorometer was used to determine emission spectrum and time profile upon excitation with UV photons. This will be compared to light profiles induced by ionizing particles in \autoref{sec:psd}.

\subsection{Light yield}
\label{sec:ly}

\begin{figure}[t!]
    \centering
    \includegraphics[width=1.00\textwidth]{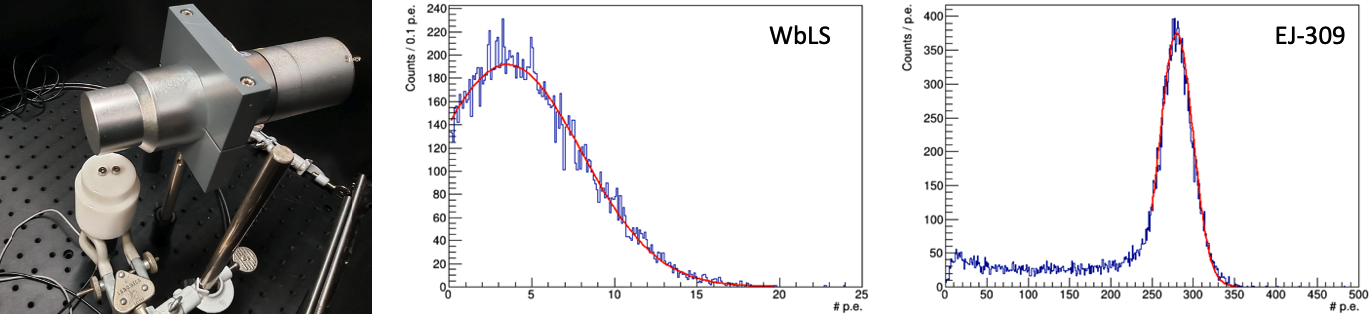}
    \caption{\textbf{Left:} LY-measurement setup with 1 inch sample cell and LaBr$_{3}$(Ce)-detector above it. \textbf{Middle:} Charge-spectrum of back-scattered $\gamma$s in WbLS. \textbf{Right:} Charge-spectrum of the reference sample EJ-309 (note: PMT signal was attenuated by -6 dB to expand the dynamic-range of the setup).}
    \label{fig:LyStudy}
\end{figure}

We study the effective light yield (i.e. scintillation+Cherenkov signals) emitted by the~WbLS based on the Compton scattering of the 662\,keV gamma rays from a $^{137}$Cs source (initial activity of 370\,kBq). The setup is shown in the left plabel of \autoref{fig:LyStudy}: The basic sample cell is a cylindrical~PTFE body of 1-inch inner diameter and height with an upper lid incorporating a round fused-silica window of 2.2\,mm thickness. This cell was completely filled with a WbLS sample and sealed under a protective nitrogen atmosphere. Within the setup, the sample cell is coupled to a Hamamatsu R9980 PMT operated at~1600\,V bias. To fix the gamma scattering angle and thus the energy deposited in the~WbLS by the recoil electron, a~$1.5\times1.5$\,inch~LaBr$_{3}$(Ce)-detector provided by~OST-Photonics is used. The setup is triggered on coincident signals in the sample cell and the~LaBr$_{3}$(Ce)-detector. The corresponding pulse spectrum in the WbLS cell is shown in the center panel of \autoref{fig:LyStudy}. Note that, absolute measurements of the amount of scintillation photons emitted for a given energy deposition in a scintillator are very difficult and prone to mismeasurements \cite{Bonhomme}. Nonetheless, to derive an absolute scale for the light yield, we do a comparative measurement with a commercial scintillator (EJ-309). The LY of this substance is stated by the supplier to reach 80$\%$ of anthracene's brightness, which produces $\sim$17400 Photons/MeV \cite{Bonhomme}. Given those numbers the LY of EJ-309 can be calculated to be $\sim$13920 Photons/MeV. The result is displayed in the right panel of \autoref{fig:LyStudy}. Note that for both result spectra, the~PMT response was calibrated to the number of detected photo-electrons by the usage of a single-photon emitting picosecond laser with a wavelength of~405\,nm. Moreover, the effect of different quantum efficiencies of the used PMT for different emission spectra of the investigated samples and also different spectral transmissions of the cell window were corrected.\\ 
By studying the reproducibility of the light yield measured for the reference sample (EJ-309, see \autoref{fig:LyStudy} right) the combined systematic uncertainties were estimated. The EJ-309 sample was measured 10 times on different days each time freshly filled into the previously cleaned empty cell. The variance of these measurements is interpreted as the influence of systematic effects like HV stability, minor temperature changes in the laboratory, efficiency of the oxygen removal from the LS and other sample preparation and cleaning related differences. By this study the relative systematic uncertainty on the LY of a given sample can be estimated to a value of 4.1\%, while statistical errors of $\sim$0.1\% were achieved for the measurements.

For the EJ-309 reference sample a mean p.e.-yield of~(280 $\pm$ 0.2)\,p.e.~(photo electrons) is observed, corresponding to the energy of the backscatter peak of 477\,keV. The width of the peak corresponds to~(19,6 $\pm$ 0.1)\,p.e.~($1\sigma$). Based on the peak p.e.~yield measured for the WbLS, i.e.~(4.5~$\pm$~0.1)\,p.e. yield, the p.e.~yield of EJ-309 and the deposited energy, the light yield of the WbLS sample can be estimated to~(223\,$\pm$\,10) photons per MeV for~$\gamma$/$e$-interactions. This result corresponds to approx.~2\% of the light yield known from purely organic~LAB-based scintillation cocktails \cite{Bonhomme} currently used by~kt-scale neutrino observatories such as ~SNO+ or~JUNO.

\FloatBarrier

\subsection{Emission spectrum} 
\label{sec:es}

The wavelength-dependent absorption and emission of our~WbLS sample was evaluated using an Edinburgh~FS5 spectro-fluorometer. A xenon lamp is used to illuminate the sample, while two gratings select the excitation and emission wavelengths recorded, each at~1\,nm resolution.

\begin{figure}[h]
    \centering
    \includegraphics[width=0.85\textwidth]{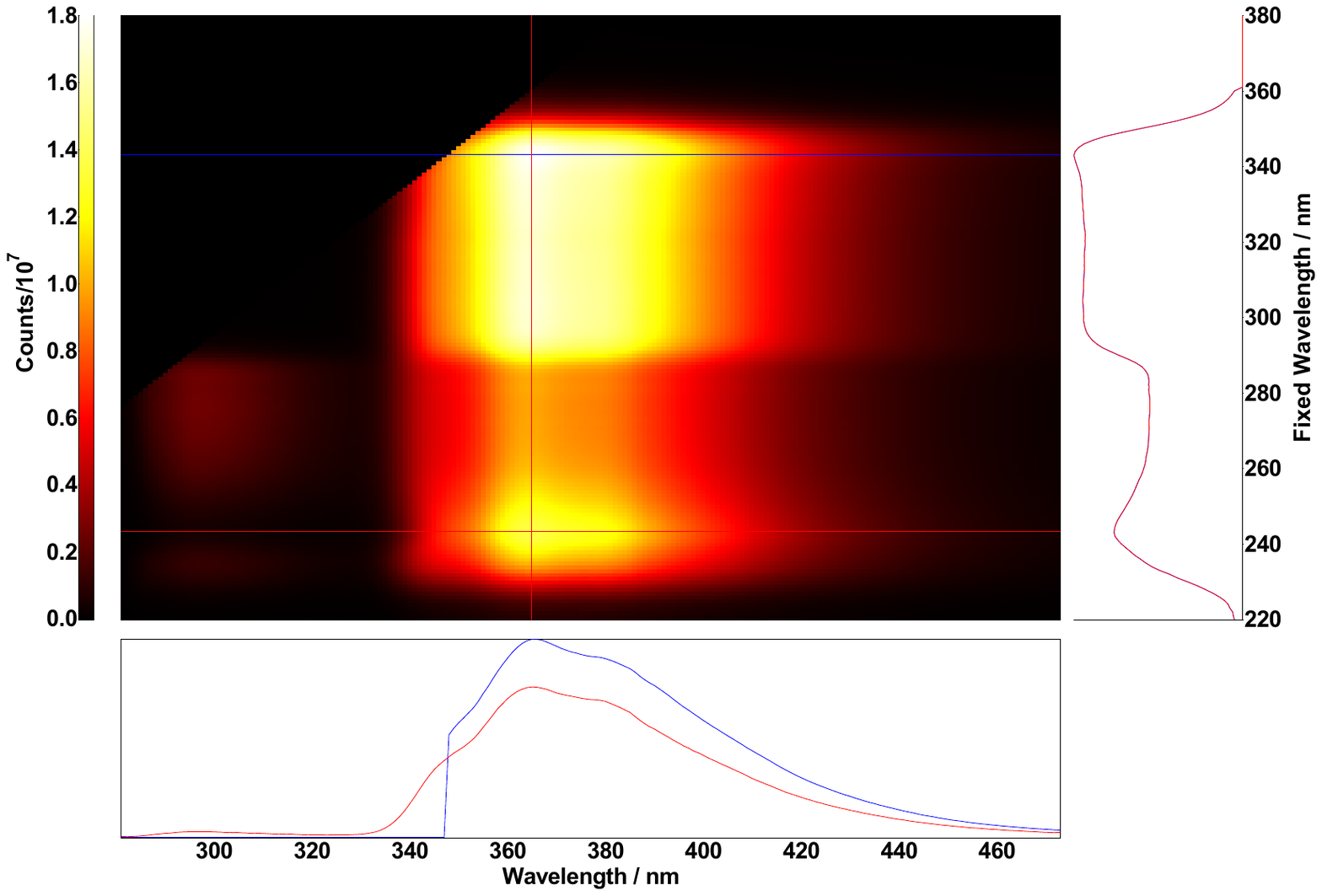}
    \caption{Absorption and emission properties of the~WbLS sample. The heat map reflects the intensity of the emitted wavelength~(x-axis) vs. the incident wavelength~(y-axis). The absorption peaks visible correspond to LAB/TX~(240\,nm) and~PPO~(340\,nm), the emission is dominated by~PPO~(370\,nm) with a weak side band from TX around~300\,nm.}
    \label{fig:absorptionemission}
\end{figure}

\autoref{fig:absorptionemission} shows a corresponding heat map of the fluorescence emission intensity plotting emission (x-axis) vs. excitation wavelength~(y-axis). This survey uses a standard cuvette with a quadratic cross-section of 10-by-10-mm$^2$ and front-face geometry, i.e.~light emission is measured on the same face of the cuvette as the incident light. There are two distinctive absorption bands: The more prominent band peaking at~340\,nm corresponds to the direct absorption of UV light by~PPO, leading to the expected re-emission band peaking at~370\,nm. The second absorption maximum at about~240\,nm is interpreted as an overlay of~LAB and~TX absorption. Apart from the prominent re-emission resulting from non-radiative transfer of the excitation to~PPO, there is a broad low-intensity side band with a maximum around~300\,nm that fits the expected TX emission~\cite{tx:2006}. Note that direct emission from LAB would peak more prominently at~280\,nm. We conclude that -- apart from the unusual side-band caused by the direct re-emission of surfactant molecules -- the WbLS emission spectrum is quite comparable to that of a conventional fully-organic scintillator. 

\begin{figure}[h]
    \centering
    \includegraphics[width=0.85\textwidth]{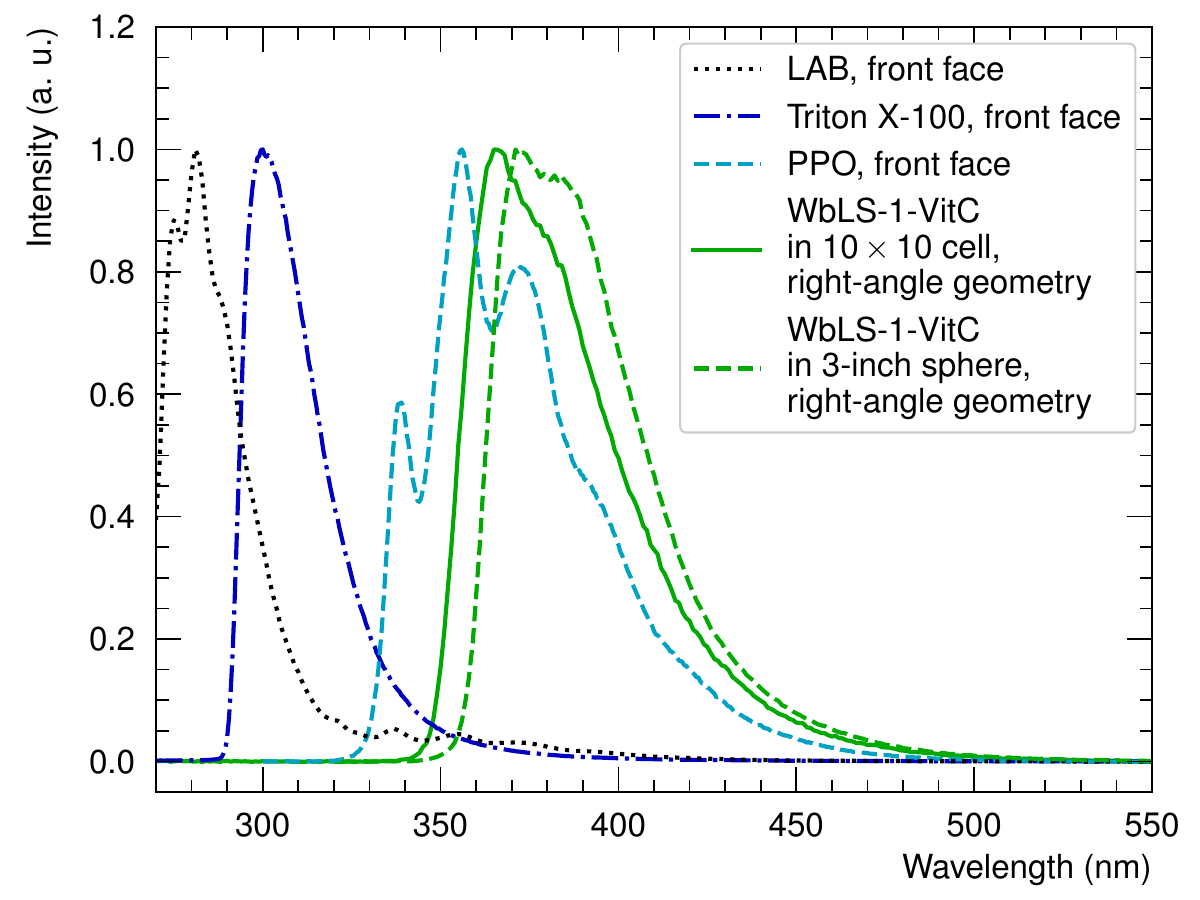}
    \caption{Emission spectra of the substances used in the sample cocktail normalized to their global maxima. To minimize the self-absorption and re-emission of the substances a front face geometry with minimal light paths through the medium was realized in the spectrometer. To study the effective light emission observable in a WbLS detector exceeding the size of a few ml a right-angle geometry with longer light paths through the sample was selected for recording the spectrum of the WbLS cocktail (see solid green line). The wavelength-shifting effect is even more visible in the larger sample in a 3-inch glass sphere. The spectrum is notably shifted towards higher wavelengths (see dashed green line).}
    \label{fig:EmSpec_WbLS-VitC}
\end{figure}

\autoref{fig:EmSpec_WbLS-VitC} illustrates the emission spectra expected for the different components of the emission spectrum. Here, a wavelength of 255\,nm was selected for the incident light beam in all cases but PPO and the 3-inch sphere where 300\,nm were used. From left to right, the emission spectra of LAB, Triton-X and PPO are displayed, with characteristic peak emissions at 280\,nm, 300\,nm and 365\,nm, respectively. In addition, the two right most (green) curves illustrate the effect of self-absorption by the bulk WbLS material in extended scintillator samples. Measuring the emitted light under an angle of $90^\circ$ so that the fluorescence light has to pass at least several millimeters of liquid, the minimum wavelength observable is shifted to higher values. This is even more pronounced for a larger sample of WbLS (in a 3-inch glass sphere used for the Legnaro measurements described in \autoref{sec:psd}). Note, however, that this effect is on a similar order of magnitude as expected for an organic scintillator.

\subsection{Fluorescence time profile (UV-excitation)} 
\label{sec:fluorescence}
\begin{figure}[h]
    \centering
    \includegraphics[width=1\textwidth]{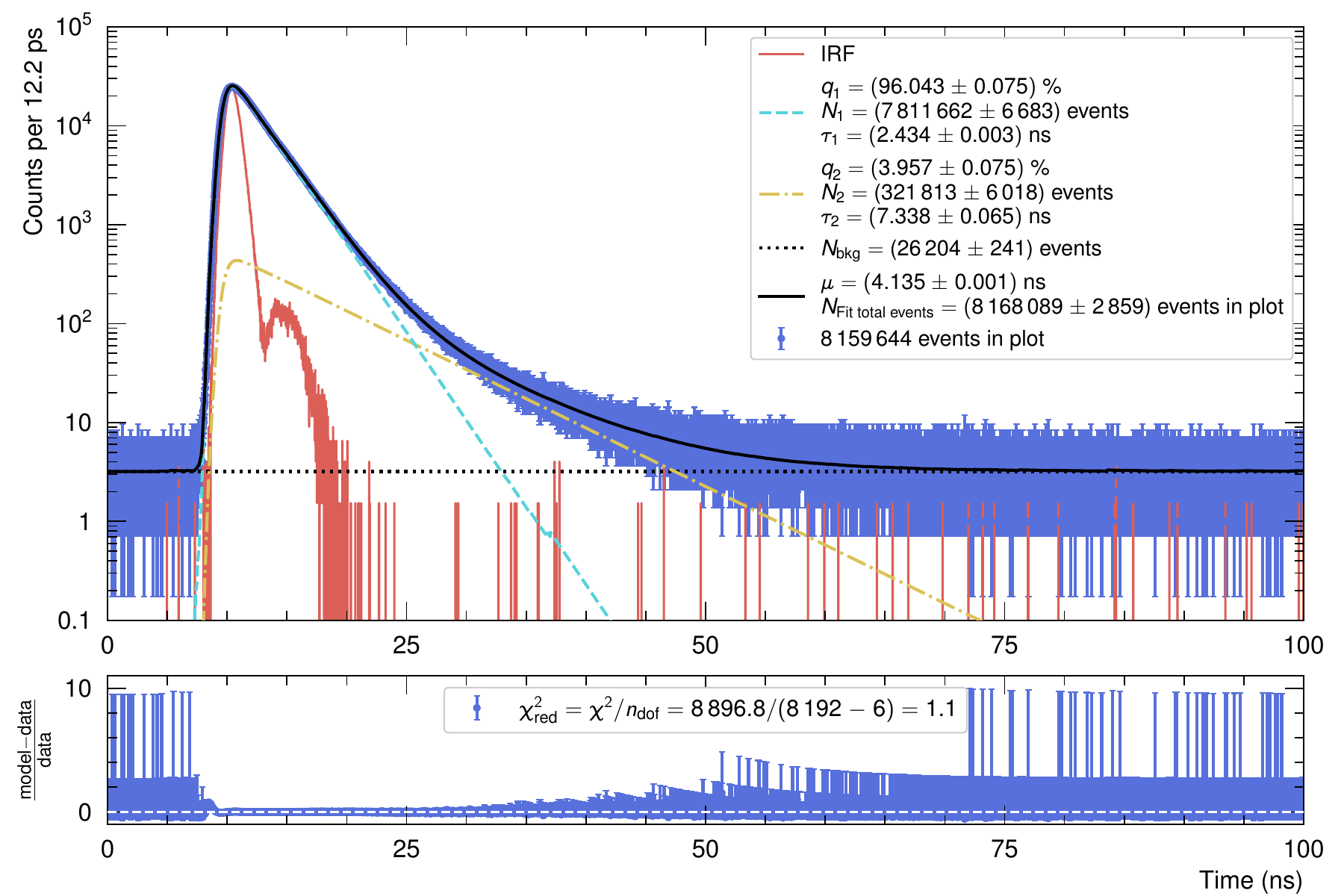}
    \caption{Fluorescence time profile for UV excitation at~255\,nm, measured in front-face geometry.}
    \label{fig:255nm_FrontFace}
\end{figure}

The fluorescence time profile for excitation with UV light was studied  with the same Edinburgh FS5 spectro-fluorometer. Here, an EPLED255 pulsed LED source was used to enable time correlated single photon counting (TCSPC). The incident wavelength ($\lambda=255\,{\rm nm}$) is mainly absorbed by LAB and TX, leading to a time response similar to the charged-particle excitation of electrons. For the measurement the sample was irradiated in front-face geometry in the corresponding holder. The instrumental response function (IRF) of the spectro-fluorometer was studied beforehand with identical measurement settings using an aqueous solution of LUDOX\textregistered HS-40 colloidal silica, that scatters promptly the laser light into the detectors of the spectrometer. The resulting IRF is shown along with the data in \autoref{fig:255nm_FrontFace}. Note, that the IRF shows a clear second population, which can be attributed to the late-pulsing population of the light detector module PMT of the FS5. 

%We have used the~FS5 spectro-fluorometer to determine as well the decay times of the fluorescence signal. For this, we have illuminated the sample with a pulsed laser and a pulsed~LED at two different wavelengths~(EPL375 and EPLED255, respectively). Direct excitation of~PPO at~375\,nm yields a relatively long fast fluorescence component of~$\tau_f=(1.65\pm0.10)$\,ns~(92\%) with a small admixture of a slow component of~$\tau_s=(6.9\pm1.0)$\,ns. Indirect excitation via the LAB (and possibly~TX) yields as expected a slightly slower fast component of~$\tau_f=(2.15\pm0.10)$\,ns. The corresponding decay fit is shown in Fig.~\ref{fig:fluorescence}, parameter values are summarized in Tab.~\ref{tab:fluorescence}. All errors quoted are systematics-dominated and where determined by varying the fit ranges.\\
%\textcolor{red}{[text has to be adjusted, show fit values in table; which graphs to show? and/or could be rearranged as 2-by-2]}

%compare with LAB +  1.5 g/l PPO, the fast component is 5.3 ns, the slower 16.7 ns, so WbLS is significantly faster.
\noindent The experimental fluorescence time profile data can be fitted well with a model based on a two-component exponential decay function $F(t)$ 

\begin{equation}\label{eq:scint_time_profileStock1}
    F(t)
        = \sum_{i=1}^{2}
            q_{i} \Theta(t) \frac{e^{-t/\tau_{i}}} {\tau_{i}},
\end{equation}
where the sum of the weights $q_{i}$ corresponds to unity. The results from the fit are summarized in \autoref{tab:fluorescence}. All uncertainties are fully dominated by systematic effects and where evaluated by varying the fit ranges.  

% \begin{figure}[h]
%     \centering
%     \includegraphics[width=0.85\textwidth]{pics/excitation_255.PNG}
%     \caption{Fluorescence time profile for UV excitation at~255\,nm. Black data points indicate instrumental response function. [FIXME: Replace by a figure with time x-axis and fit parameters displayed in legend.]}
%     \label{fig:fluorescence}
% \end{figure}

\begin{table}[h]
    \centering
    \begin{tabular}{l|cc}
    \hline
    Decay Profile    & $\tau_i$ [ns] & $q_i$ [\%]  \\
    \hline
    fast component 1 & $2.434\pm0.003$ &  $96.04\pm0.08$ \\ 
    slow component 2 & $7.338\pm0.007$ & $3.96\pm0.08$\\
    \hline
     \end{tabular}
    \caption{The measured scintillation emission timing parameters for excitation with UV light with a wavelength of 255\,nm. Note, that this light emission is particularly fast, as purely organic scintillation mixtures such as LAB + 1.5 g/l PPO show a fastest decay component of about $\sim$5.3\,ns, while the second slower one is $\sim$16.7~ns.}
    \label{tab:fluorescence}
\end{table}

% \begin{figure}[h]
%     \centering
%     \includegraphics[width=0.7\textwidth]{pics/300nm_300nm_ConvIRF_Exp_N1N2_Nbkgfixed}
%     \caption{300 nm front face}
%     \label{fig:300nm_FrontFace}
% \end{figure}

% \begin{figure}[h]
%     \centering
%     \includegraphics[width=0.7\textwidth]{pics/375nm_375nm_ConvIRF_Exp_N1N2_Nbkgfixed}
%     \caption{375 nm front face PileUp? \url{https://www.picoquant.com/images/uploads/page/files/7253/technote_tcspc.pdf}}
%     \label{fig:375nm_FrontFace}
% \end{figure}

% \begin{figure}[h]
%     \centering
%     \includegraphics[width=0.7\textwidth]{pics/WbLS1VitC_SG_255nm_ConvIRF_RiseExp_N1N2_Nbkgfixed}
%     \caption{255 nm right-angle geometry. Fit model with correction term \url{https://arxiv.org/ftp/arxiv/papers/1711/1711.01137.pdf}
%     Pile up correction? Here Self-absorption correction?}
%     \label{fig:255nm_SG}
% \end{figure}

\FloatBarrier

\subsection{Scintillation profile (particle excitation) and Pulse Shape Discrimination}
\label{sec:psd}

One of the most powerful background suppression techniques in LS detectors is Pulse Shape Discrimination (PSD). Based on the type of ionizing particle, more precisely its differential energy loss {d$E/$d$x$} in the LS material, there is a difference in the absolute light output (quenching) and crucially in the observed fluorescence time profiles. For a theoretical description and further empirical details see Ref.~\cite{Foerster, Birks, Horrocks}. The~PSD performance depends on the magnitude of the difference in the emission time profiles for different particle species. Here, we used signals induced by neutrons (proton recoils in the WbLS) and gamma radiation (electron recoils) to quantify the PSD performance.

\begin{figure}[t!]
    \centering
    \includegraphics[width=1.0\textwidth]{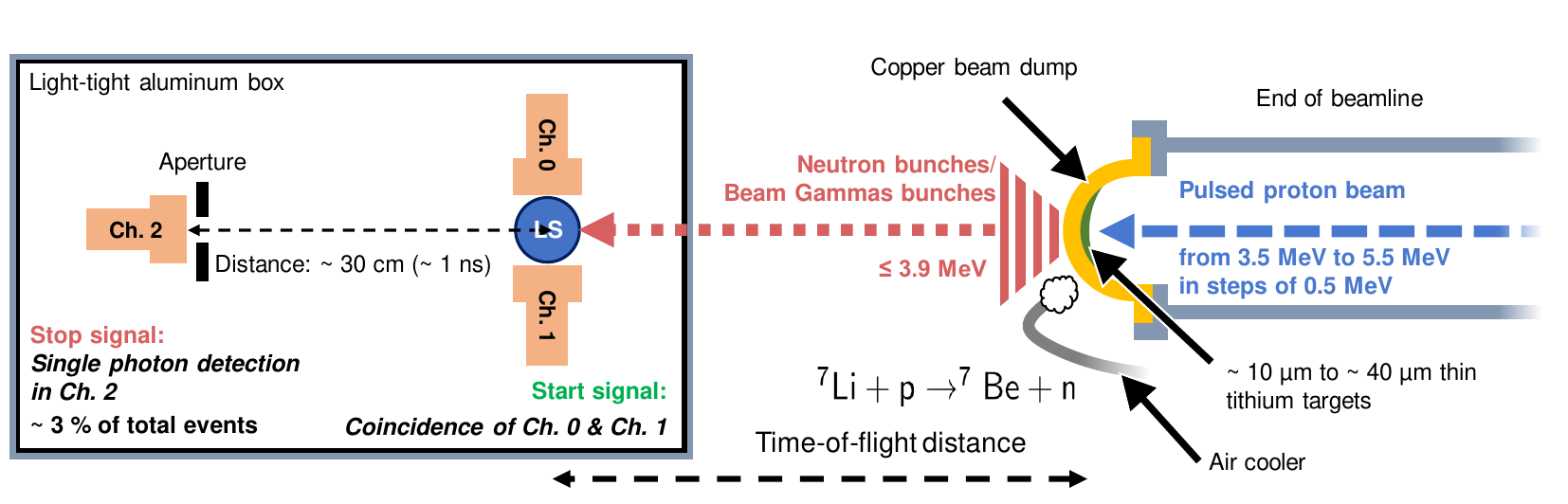}
    \caption{Illustration of the scintillation time profile experiment at the end of the beamline of the~CN Van de Graaff accelerator at~INFN Legnaro. The experimental setup exploits the~TCSPC technique. The start signal of the time measurement is gained from the coincident pulsing of two PMTs directly on the WbLS vessel. A stop signal is generated based on a single photon hit acquired by a~PMT placed at a distance of~60\,cm from the LS cell. Neutrons are generated by a proton beam impinging on a thin metallic~Li target \cite{HansPhD}.}
    \label{fig:FluorSetup}
\end{figure}

\paragraph{Experimental Setup.} For the present study, we were able to reuse the experimental setup shown in~\autoref{fig:FluorSetup}. It was originally developed for a detailed characterization of the PSD capabilities of the JUNO organic liquid scintillator \cite{HansPhD, RaphaelPhD}. Neutron and gamma irradiation is obtained with the help of a pulsed low-energy (3.5\,$-$\,5.5\,MeV) proton beam impinging on a  thin metallic lithium target. In the corresponding reaction, neutrons with energies from 1.86 to 3.86\,MeV as well as gamma rays are produced. The scintillator sample is placed 1.5\,m from the reaction point to allow signal separation based on the neutron time-of-fligt delay.
%Signal categories are it was placed at the end of the~0$^{\circ}$ beamline of the~CN accelerator at~INFN Legnaro. Here, pulsed proton beams with energies from~3.5\,MeV up to ~5.5\,MeV~($\sigma_{E}\sim$~3\,keV) can be guided onto a thin metallic lithium target~(typ.~5-20\,$\upmu$m). In the nuclear reaction 

%\begin{equation}
%\text{p} +~^{7}\text{Li} \longrightarrow~^{7}\text{Be} + \text{n}
%\end{equation}

%quasi-monoenergetic neutrons~(QMN) adjustable between~3.86\,MeV down to~1.86\,MeV as well as beam correlated gammas can be created. The detector setup is placed in a distance of~1.5\,m from the lithium target. The bunch widths of typ. well below~1\,ns and a repetition rate of~600\,kHz allow time-of-flight~(ToF) discrimination of neutrons and gammas.\\
As above, the light emission profile of the WbLS sample by time correlated single photon counting (TCSPC), similar to \cite{Bollinger, Marrodan}. The sample was placed in a spherical borosilicate glass vessel (diameter of~$\sim$72\,mm, wall thickness $<$1\,mm) and placed between two ~ETEL~9821B~PMTs (2.75'') that provide the start signal, while a third identical PMT at 60\,cm distance receives only single photons and provides the stop. PMT traces with a total length of~1000\,ns were digitized by a high-performance flash~ADC (Agilent Acqiris,~10\,bit,~2\,GS/s). The IRF was determined with a picosecond pulsed diode laser system (NKT Photonics). See Refs.~\cite{HansPhD, StockProceed} for a more detailed description of the experimental setup.

%The detector setup for the fluorescence time profile measurement is placed onto optical benches in a darkbox~(wall thickness~1\,mm) made of aluminum that is also acting as a faraday cage. A spherical borosilicate glass vessel with an outer diameter of~$\sim$72\,mm and wall thickness below~1\,mm contains a sample of~$\sim$150\,ml~LS. To prevent oxidation of the LS, the remaining volume in the sphere is filled with a protective nitrogen atmosphere with an overpressure of some millibar. The vessel is enclosed by two gas-tight stopcocks with~PTFE plugs. The sphere with the~LS is placed between two~ETEL~9821B~PMTs with a~68\,mm photocathode~\mbox{\cite{HansPhD, StockProceed}}. 

Based on the time structure of the beam, neutrons and~$\gamma$s were distinguished by~ToF, realizing a purity of the neutron sample above~99.8\%. To enhance the electron-like event sample, the relatively low yield of the Li-target was enhanced by including data samples taken with a $^{137}$Cs source ($A=370$\,kBq). The corresponding data samples contain $4.3\times10^5$ ($8.7\times10^5$) electron (proton) recoil events.

\begin{figure}[b!]
    \centering
    \includegraphics[width=0.85\textwidth]{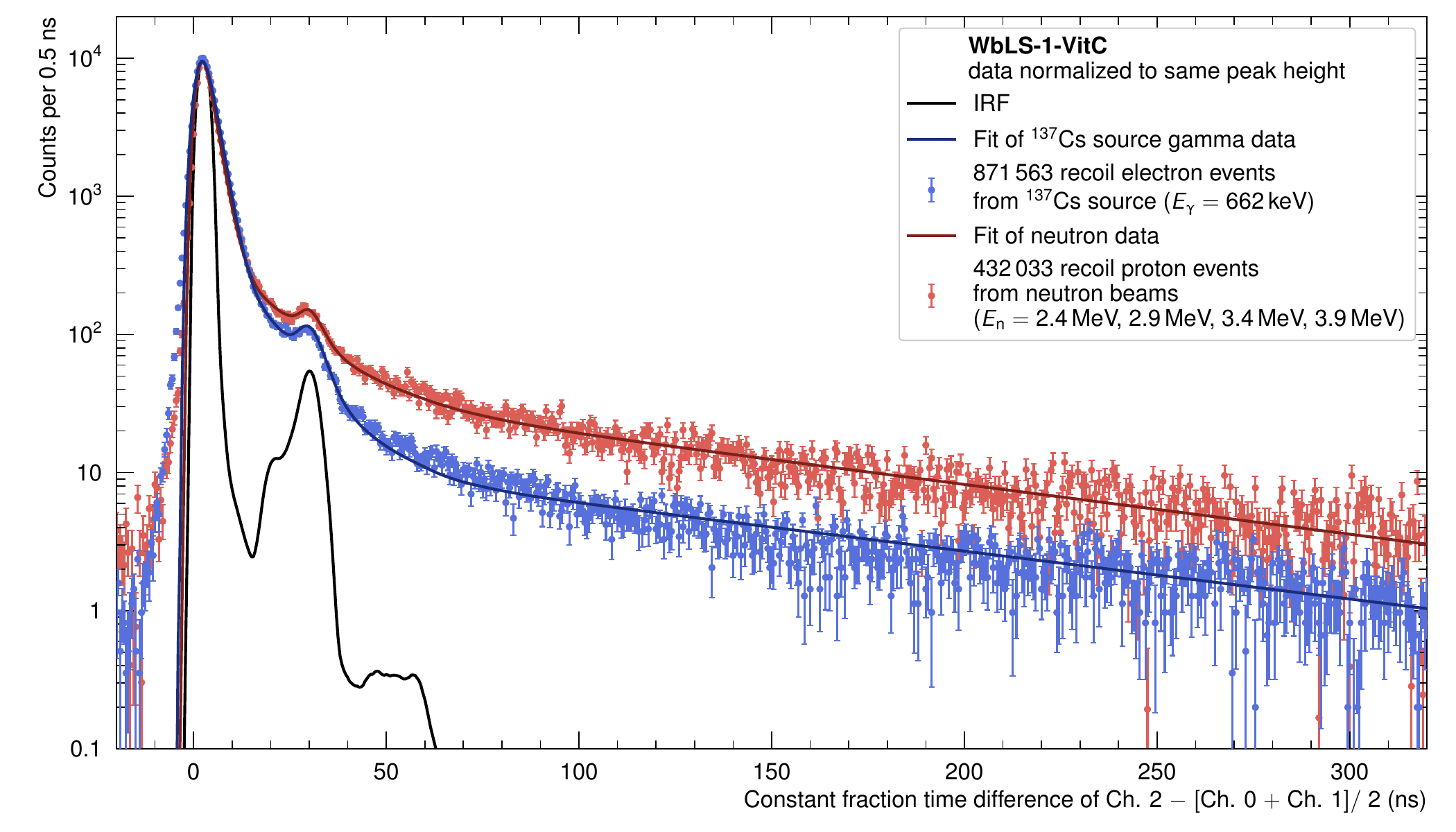}
    \caption{Electron recoil (blue) and proton recoil (red) time profiles of the WbLS sample. The black solid line indicates the IRF of the setup, governed by the single photon time response of Channel 2.}
    \label{fig:FluorWbLS-VitC}
\end{figure}

\paragraph{Time profile.} The model $S(t)$ chosen to describe the scintillation pulses contains three exponential decay functions (components $\tau_i,n_i$) and a rise time parameter $\tau_r$,
The scintillation time profile $S(t)$ is modeled as 

\begin{equation}\label{eq:scint_time_profileStock1}
    S(t)
        = \sum_{i=1}^{3}
            q_{i} \Theta(t) \frac{e^{-t/\tau_{i}} - e^{-t/\tau_{r}}}{\tau_{i} - \tau_{r}},
\end{equation}
with the sum of the weights $\sum_iq_i=1$ and the average scintilaltion life time $\tau_{\rm life}$ the weighted sum of the individual $\tau_i$ ,
\begin{equation}\label{eq:EffectLifeStock}
    \tau_{\text{life}}
        = \sum_{i=1}^{3}
            q_{i} \tau_{i} + \tau_r.
\end{equation}
%using the weighted sum over the individual lifetime components in the respective fit. 
Since here a different data parameterization than in~\cite{Lifetime} is applied, the scintillation rise time can be included. This emission profile is convoluted with the IRF of the PMT system determined with the picosecond laser.
%\enlargethispage{1.0cm}  
%As mentioned in~\ref{ExpSetupSubs} the~IRF is measured by means of a picosecond laser. The time response of the setup is governed by the transit time spectrum for single photon hits of the distant PMT~(Ch.~2). To also account for the smearing of the time profile start signal, which is partially caused by the~WbLS sample but mainly by the time jitter of the used~PMTs, the in-situ gained data for this coincidence is convoluted. The resulting~IRF is drawn as the black solid line in \autoref{fig:FluorWbLS-VitC} and shows next to the nearly Gaussian main peak the expected distinct late pulsing population. The fit model is thereby the numerical binwise convolution of the~IRF with the three exponential decays including the rise time and an additional constant background component. 
The results of all fits applied to the data for both, neutron and gamma interactions, are given in \autoref{tab:WbLS1VitC_RecoilElectronProtonFitResults}. As in the case of UV excitation, the observed fluorescence time profiles are very fast, with a shorter component $\tau_1$ in the order of 2\,ns. This result is comparable to earlier measurement of WbLS emission profiles reported in Refs.~\cite{Callaghan:2022ahi,Callaghan:2023oyu}.

\begin{table}[t!]
	\centering
	\begin{tabular}{ccc}
		\hline
		\hline		
		\noalign{\vskip 0.25em}
		& \multicolumn{2}{c}{\textbf{WbLS-1-VitC}} \\
		\noalign{\vskip 0.25em}
		Source & e$^{-}$ ($^{137}$Cs source) & p (neutron beams) \\
		\noalign{\vskip 0.25em}
		\hline
		\hline
		\noalign{\vskip 0.25em}
		$q_1$ (\%) & $88.288 \pm	0.134 \,^{+1.071}_{-0.138}$ &  $82.312 \pm 0.178 \,^{}_{-1.218}$ \\
		\noalign{\vskip 0.25em}
		$q_2$ (\%) & $9.216 \pm 0.117 \,^{+0.175}_{-1.303}$ &  $9.464 \pm 0.117 \,^{+0.764}_{}$ \\
		\noalign{\vskip 0.25em}
		$q_3$ (\%) &  $2.496 \pm 0.033 \,^{+0.223}_{}$ &  $8.224 \pm 0.100 \,^{+0.793}_{}$ \\
		%$q_4$ (\%) & $\pm \,^{+}_{-}$ &  $\pm \,^{+}_{-}$ & $\pm \,^{+}_{-}$ &  $\pm \,^{+}_{-}$ \\
		\noalign{\vskip 0.25em}
		\hline
		\noalign{\vskip 0.25em}
		$\tau_{\text{r}}$ (ns) &  $1.501 \pm 0.016 \,^{+0.352}_{-0.277}$ &  $0.745 \pm 0.008 \pm \,^{+0.420}_{-0.419}$ \\
		\noalign{\vskip 0.25em}
		\hline
		\noalign{\vskip 0.25em}
		$\tau_1$ (ns) & $2.072 \pm 0.021	\,^{+0.012}_{-0.047}$ &  $2.334	\pm 0.024 \,^{}_{-0.262}$ \\
		\noalign{\vskip 0.25em}
		$\tau_2$ (ns) & $10.305 \pm 0.104 \,^{+1.140}_{}$ &  $14.727 \pm 0.148 \,^{}_{-3.008}$ \\
		\noalign{\vskip 0.25em}
		$\tau_3$ (ns) & $125.294	\pm 2.011 \,^{+40.689}_{}$ & $122.396	\pm 1.974 \,^{}_{-18.003}$ \\
		%$\tau_4$ (ns) &	$\pm \,^{+}_{-}$ &  $\pm \,^{+}_{-}$ & $\pm \,^{+}_{-}$ &  $\pm \,^{+}_{-}$ \\
		\noalign{\vskip 0.25em}
		\hline
		\noalign{\vskip 0.25em}
		$\tau_{\text{life}}$ (ns) & 7.408 $\pm 0.072\,^{+1.116}_{-0.311}$ & $14.126	
		\pm 0.206 \,^{+1.064}_{-1.580}$ \\
		\noalign{\vskip 0.25em} 
		\hline
		%	\hline		
	\end{tabular}
	\caption{Parameters of the recoil electron and recoil proton time profile models for the water-based liquid scntillator~WbLS-1-VitC. Provided values include statistical and asymmetric systematic uncertainties.}
	\label{tab:WbLS1VitC_RecoilElectronProtonFitResults}%Values Cheched q is 100
\end{table}

\paragraph{PSD capability} is evaluated based on the tail-to-total method which compares the integrated scintillation tail with the total light emission of the scintillation pulses. We compare the results obtained with the WbLS sample to those of standard LAB- and pseudocumene-based scintillators with varying PPO content that were studied in the same setup. For this, the fit functions describing the acquired scintillation signals are time-aligned at their maximum probability. For each sample, the optimum tail-to-total ratio is determined by varying the start time of the tail integration and obtaining the maximum tail-to-total difference $\Delta \mu$ for neutrons and gammas. Results are quoted in \autoref{tab:PSD_evaluation}. From the comparison, we conclude that the WbLS samples performs surprisingly well, comparable to an LAB-based scintillator with 1.5\,g/l PPO.

%t the pulse shape discrimination capabilities of the WbLS with conventional LAB- and pseudocumene-based liquid scintillator mixtures the tail-to-total method, which compares the integrated scintillation tail with the total light emission from the beginning to the end of a scintillation event, is applied. Therefore, the time profile models~(probability density functions), derived from the corresponding fits, are shifted such that their maxima occur at the same position in time. The integration windows for the tail ranging from different start times to~500\,ns after the peak of the time profile. The difference of the areas below the curves (neutron area minus gamma area), is the tail-to-total difference~$\Delta\mu$. The start time of the tail integration after the peak is thereby optimized such, that~$\Delta\mu$ reaches its maximal value.\\
%When comparing the~LAB/PPO-based scintillators with WbLS, a similar PSD performance can be observed with the sample containing even 1.5 g/l PPO. Note, that the PPO content in WbLS is 1 g/l. Nonetheless, the pulse shape difference of a Borexino-like LS based on pseudocumene which is considered to be the largest in monolithic neutrino detectors to date, is not reached (see \autoref{tab:PSD_evaluation}). 

\begin{table}[b!]
    \centering
    \begin{tabular}{|c|c|c|}
         \hline 
         Sample & Maximal tail-to-total difference $\Delta \mu$ & Time after time profile peak (ns) \\ \hline 
         WbLS-1-VitC & $0.070 \pm 0.010$ & $9.67 \pm 1.92$ \\ \hline 
         LAB + 0.5\,g/l PPO & $0.038 \pm 0.001$ & $44.33 \pm 4.83$ \\ %\hline
         LAB + 1.5\,g/l PPO & $0.067 \pm 0.001$ & $16.33 \pm 1.29$ \\ %\hline 
         LAB + 2.0\,g/l PPO & $0.106 \pm 0.004$ & $11.33 \pm 1.65$ \\ \hline 
         %PC + 1.5\,g/l PPO & 0.168 & 7.7 \\ \hline 
         PC + 1.5\,g/l PPO & $0.166\pm 0.002$ & $7.83 \pm 0.22$ \\ \hline 
    \end{tabular}
    \caption{Comparison of the tail-to-total difference between recoil proton and recoil electron events. The start time for the tail integration after the time profile peak was selected such, that it maximizes~$\Delta\mu$.}
    \label{tab:PSD_evaluation}
\end{table}

\section{Conclusions}
\label{sec:conclusions}

The present paper describes a novel formulation of Water-based Liquid Scintillator (WbLS) based on LAB as the solvent and Triton-X as surfactant. As shown in \autoref{sec:transparency}, the production recipe described in \autoref{sec:production} leads to the creation of micelles with an average diameter of only 2.8\,nm, causing a negligible amount of Rayleigh scattering in our sample. The measured attenuation length of 6\,m at 430\,nm is likely fully dominated by the transparency of the distilled water used for the WbLS production.

The estimated light yield of (223\,$\pm$\,10) photons per MeV for electron-like signals is quite high given that the relative fraction of only 1\% of LAB in the WbLS composition (\autoref{sec:ly}). However, Triton-X is believed to participate in the scintillation/energy transfer of the micelles. Some hint for this is provided by the observation of a side band in the 2D fluorescence profile (\autoref{fig:absorptionemission}) that can be attributed to the direct light emission of Triton-X. Note the relatively large amount of 13\% weight in the WbLS cocktail.

Finally, we have investigated the fluorescence time profile of the WbLS, both under UV and particle excitation. As observed for other WbLS recipes \cite{Callaghan:2022ahi,Callaghan:2023oyu}, time emission is particularly fast ($\tau_1=2.1-2.4$\,ns, \autoref{sec:fluorescence}) especially when compared to conventional organic scintillators. This illustrates the necessity of fast photo sensors like LAPPDs \cite{Adams:2013nva,Kaptanoglu:2021prv} when using WbLS as a hybrid medium for Cherenkov/scintillation detection. Based on the signals of neutrons and gammas acquired with our WbLS sample (\autoref{sec:psd}), we investigated as well the Pulse Shape Discrimination capabilities for neutrons and gammas. We find a performance comparable to an organic LAB-based scintillator with 1.5\,g/l, suggesting that PSD can indeed be used for particle identification in future experiments, complementing the discrimination capabilities resulting from the Cherenkov/scintillation ratio (e.g.~Ref.~\cite{Sawatzki:2020mpb,Zsoldos:2022mre}).

To our knowledge, the present paper provides the most complete description of a particular WbLS sample currently available. We hope that it can serve as a reference for the further development of the technology and as input for WbLS samples used in simulation studies of a future large-scale neutrino (or dark matter) experiment employing WbLS as a target (veto) medium. 

\FloatBarrier

\acknowledgments

This work benefited substantially from the support and funding by the Detector Laboratory of the Cluster of Excellence~PRISMA$^+$. We are very grateful to the staff of the Detector Division around Dr. Quirin Weitzel. Our special thanks go to the chemotechnical employee of the Mainz Institute for Physics Joachim Strübig. \\
Moreover, we are very grateful for the support of the Laboratori Nazionali di Legnaro~(LNL) of the Italian Istituto Nazionale di Fisica Nucleare~(INFN). For the excellent collaboration during our beamtimes we especially thank Dr. Pierfrancsco Mastinu and Dr. Elizabeth Musacchio as well as the two operators of the CN accelerator Luca Maran and Daniele Lideo.\\ 
We are especially grateful to Ed Callaghan (UCB $\&$ LBNL) for his generous help during processing and handling the beamtime data on the Savio computing cluster at Berkeley.\\ 
For countless detailed and inspiring discussions we would like to thank especially Prof. Dr. Franz von Feilitzsch\mbox~(TUM), Prof. Dr. Garbriel Orebi Gann (UCB $\&$ LBNL), Dr. Minfang Yeh (BNL),  Dr. Brennan Hackett~(Max Planck Institute for Physics) and Andreas Leonhardt~(TUM).\\  
%Likewise, special thanks to Prof. Dr. Gabriel Orebi-Gann's group at the University of California at Berkeley (UCB) and Lawrence Berkeley National Laboratory (LBNL). The numerous discussions, loans of equipment, and supportng of our beamtimes during the past three years contributed greatly to the success of our measurements.\\  
The development of this novel scintillation medium was supported by the~BMBF Collaborative Project~05H2018~-~R\&D Detectors~(Scintillator). Moreover, this research was co-funded by the Cluster of Excellence ORIGINS which is funded by the Deutsche Forschungsgemeinschaft~(DFG, German Research Foundation) under Germany’s Excellence Strategy~–~EXC-2094~–~390783311.

%This work benefited substantially from the support by the Detector Laboratory of the Cluster of Excellence~PRISMA$^+$. We are very grateful to the staff of the Detector Division around Dr. Quirin Weitzel.\\
%The development of this novel scintillation medium was funded by the~BMBF Collaborative Project 05H2018~-~R\&D Detectors~(Scintillator). 

%\paragraph{Note added.} This is also a good position for notes added
% \vspace{2cm}

% Citation for Triton-X properties:\\
% Zhao J, Wei YJ. [Fluorescence spectra and fluorescence quantum yield of triton X-100]. Guang Pu Xue Yu Guang Pu Fen Xi. 2006 Aug;26(8):1523-5. Chinese. PMID: 17058962.\\
% Abstract:\\
% Fluorescence spectra and fluorescence quantum yield of triton X-100 (TX) aqueous solution are reported. In strong acidic solutions, TX gives no fluorescence. When pH > 1, TX gives a strong and steady fluorescence with the maximum excitation wavelengths at 229 and 275 nm, and the maximum emission wavelength at 302 nm, respectively. TX aqueous solution can produce resonance fluorescence. The resonance fluorescence peak is located at 285 nm. A linear relationship between the fluorescence intensity and TX concentration was found in the range of 0. 1-90 mg x L(-1). The detection limit of TX is 0.1 mg x L(-1). By using L-tryptophan as a reference, the fluorescence quantum yield of TX at maximum excitation wavelength 280 nm was measured to be 0.121.


\bibliographystyle{JHEP.bst}
\bibliography{main.bib}

% \begin{thebibliography}{99}

% %%%%%%%%%
% % BIBTEX
% %%%%%%%%

% \bibitem{a}
% Author, \emph{Title}, \emph{J. Abbrev.} {\bf vol} (year) pg.

% \bibitem{b}
% Author, \emph{Title},
% arxiv:1234.5678.

% \bibitem{c}
% Author, \emph{Title},
% Publisher (year).

% % Please avoid comments such as "For a review'', "For some examples",
% % "and references therein" or move them in the text. In general,
% % please leave only references in the bibliography and move all
% % accessory text in footnotes.

% % Also, please have only one work for each \bibitem.

% \end{thebibliography}
\end{document}